# Potential mechanisms for imperfect synchronization in parkinsonian basal ganglia

Choongseok Park[1], Leonid L. Rubchinsky[1,2]

[1] Department of Mathematical Sciences and Center for Mathematical Biosciences, Indiana University Purdue University Indianapolis, Indianapolis, IN 46202, USA
[2] Stark Neurosciences Research Institute, Indiana University School of Medicine, Indianapolis, IN 46202, USA

## Abstract

Neural activity in the brain of parkinsonian patients is characterized by the intermittently synchronized oscillatory dynamics. This imperfect synchronization, observed in the beta frequency band, is believed to be related to the hypokinetic motor symptoms of the disorder. Our study explores potential mechanisms behind this intermittent synchrony. We study the response of a bursting pallidal neuron to different patterns of synaptic input from subthalamic nucleus (STN) neuron. We show how external globus pallidus (GPe) neuron is sensitive to the phase of the input from the STN cell and can exhibit intermittent phase-locking with the input in the beta band. The temporal properties of this intermittent phase-locking show similarities to the intermittent synchronization observed in experiments. We also study the synchronization of GPe cells to synaptic input from the STN cell with dependence on the dopamine-modulated parameters. Earlier studies showed how the strengthening of dopamine-modulated coupling may lead to transitions from non-synchronized to partially synchronized dynamics, typical in Parkinson's disease. However, dopamine also affects the cellular properties of neurons. We show how the changes in firing patterns of STN neuron due to the lack of dopamine may lead to transition from a lower to a higher coherent state, roughly matching the synchrony levels observed in basal ganglia in normal and parkinsonian states. The intermittent nature of the neural beta band synchrony in Parkinson's disease is achieved in the model due to the interplay of the timing of STN input to pallidum and pallidal neuronal dynamics, resulting in sensitivity of pallidal output to the phase of the arriving STN input. Thus the mechanism considered here (the change in firing pattern of subthalamic neurons through the dopamine-induced change of membrane properties) may be one of the potential mechanisms responsible for the generation of the intermittent synchronization observed in Parkinson's disease.

## Introduction

A variety of oscillatory rhythms are widely observed in various neural systems from central pattern generators to human central nervous system [1]. Complex interactions within a nucleus or between nuclei, which are often accompanied by synchronization of the involved cells, are suggested to be responsible for such rhythms. Synchronized oscillations have been extensively studied because of their correlations with both normal functioning of the nervous system and abnormal states such as Parkinson's disease, schizophrenia and epilepsy [2,3].

One important example of synchronized oscillations can be found in Parkinson's disease (PD), a neurodegenerative disorder resulted from the depletion of dopamine within basal ganglia (BG). Experimental results have demonstrated strong connections between synchronous oscillations of neural activity within BG in the beta band (loosely defined as 10-30 Hz) and the hypokinetic symptoms of PD such as bradykinesia (slowness of movement) and rigidity (stiffness of joints) [4,5,6,7,8]. We showed that the synchronization of neuronal activity in PD is imperfect; the degree of phase-locking fluctuates in time in a specific manner: phase-locking is interrupted by frequent, yet short desynchronization events [9]. Interestingly, synchronization in the other parkinsonian symptom (tremor) also fluctuates in time [10,11]. Considering that synchrony may be even weaker and less stable in the healthy state, this matches well with the idea that transient dynamics in the nervous system is ubiquitous and is an essential feature for physiological functions [12].

Thus, there is a substantial importance in studying how synchronous oscillatory dynamics in Parkinson's disease may be supported by the interaction of two nuclei, external globus pallidus (GPe) and the subthalamic nucleus (STN) [13,14]. Modeling work has shown that the STN-GPe network can generate the characteristic temporal structure of oscillatory activity in the beta band similar to that observed in experiments [15]. The relatively broad range of parameters supporting this dynamics suggests that this activity pattern is a generic feature of the STN-GPe network. This property is probably grounded in the intrinsic cellular properties of the cells as well as the network architecture.

Best et al. [16] demonstrated the ability of a GPe cell to generate an irregular response to a periodic single pulse input from an STN cell. They first studied the response of a GPe cell to a brief impulse from the STN cell and constructed a phase response curve (PRC). Using this PRC, they derived a one-dimensional map when the GPe cell is periodically perturbed and showed how the generation of complex, irregular firing patterns depend on several parameters such as the bifurcation structure of the GPe cell and the speeds of the GPe cell's slow variable [$Ca$] during the active and silent phases. Considering these facts, one of the important factors in the generation of intermittent synchrony may be the response pattern of the GPe cell to beta rhythmic input due to its bursting structure. The PRC of a bursting cell is relatively simple if the duration of input is sufficiently long [17]. However, if the input duration is short and comparable to the duration of the active phase of the GPe cell as in [15], then phase sensitivity depending on timing of perturbation and the duration of synaptic input develops and produces qualitatively different entrainment of the GPe cell under rhythmic input.

In this study, given the experimentally observed intermittent synchronous dynamics [9], we consider the response of a GPe cell under periodic input from an STN cell and show that the GPe cell can generate an irregular response to beta rhythmic STN input and that the temporal structure of synchrony between the input and the output shows intermittent patterns. The effect of dopamine-modulated synaptic strength (such as the strength of the GPe-STN projection) on the network activity patterns was studied previously [15,16,18,19]. However, dopamine affects not only synaptic properties, but also cellular properties. In particular, alterations in dopamine levels lead to changes in the temporal features of firing patterns of STN neurons [20,21]. In the current study, we considered how dopamine-dependent features of STN neuronal firing modulate the GPe response and the synchrony between STN input and GPe output. This simple network was able to demonstrate the overall transition from a low coherence state to a high coherence state through intermittent synchronous activity, as the STN activity pattern (influenced by changes in the STN properties modulated by dopamine) varies. These results support the important role of GPe cells within the BG circuit and demonstrate that the response of the GPe cell, which was achieved through the interaction between the intrinsic properties of GPe cells themselves and firing properties of STN cells, provide a logical explanation for the temporal characteristics of synchrony observed in experimental settings.

## Methods

### 1. Neuronal Model

This study considers the dynamics of a model GPe cell under external excitatory synaptic input coming from STN. While this is clearly a substantial simplification of complex GPe-STN interactions even in the modeling perspective [13,15,16,19], it allows one to focus on the effect of the STN firing properties on the synchrony between STN and GPe. To represent STN input to GPe, we use the activity patterns generated by a model as well as experimental STN activity recorded in vivo in humans.

The experimental STN data were obtained in intraoperative recordings from the STN of parkinsonian patients. This data is the same as used in [9]. The extracellular signal was recorded through a microelectrode, spikes were detected in a standard manner and sorted for single units. The recordings were done in accordance with a standard surgical protocol (see [9] for details). These data were recorded in patients with advanced Parkinson's disease and we use them primarily for illustrative purposes, while model STN data are used extensively because one can vary the model parameters in an a desired way.

For GPe neuron, we used the conductance-based one-compartment model utilized in [15,16,19]. The model includes a leak current ($I_L$), fast spike-producing potassium and sodium currents ($I_K$ and $I_{Na}$), low threshold T-type $Ca^{2+}$-current ($I_T$), high-threshold $Ca^{2+}$-current ($I_{Ca}$), and $Ca^{2+}$- activated voltage-independent afterhyperpolarization $K^+$-current ($I_{AHP}$), which are responsible for the oscillatory properties of the cell. The membrane potential, $V$, obeys the following differential equation:

$$C\frac{dV}{dt} = -I_L - I_K - I_{Na} - I_T - I_{Ca} - I_{AHP} - I_{syn} + I_{app} \quad (1)$$

where $I_L = g_L(V - V_L)$, $I_K = g_K n^4 (V - V_K)$, $I_{Na} = g_{Na} m_\infty^3(V) h (V - V_{Na})$, $I_T = g_T a_\infty^3(V) r (V - V_{Ca})$, $I_{Ca} = g_{Ca} s_\infty^2(V)(V - V_{Ca})$, and $I_{AHP} = g_{AHP}([Ca]/([Ca] + k_1))(V - V_K)$. $[Ca]$ is the concentration of intracellular $Ca^{2+}$ ions, and the equation of the calcium balance is

$$d[Ca]/dt = \varepsilon(-I_{Ca} - I_T - k_{Ca}[Ca]) \quad (2)$$

Gating variables $n$, $h$ and $r$ are described by the first-order kinetic equations of this type:

$$dx/dt = \phi_x(x_\infty(V) - x)/\tau_x(V) \quad (3)$$

$I_{app}$ is a constant applied current (which may be thought of as a parameter, which includes striatal input, approximated as a constant). Voltage dependent activation and inactivation variables and time constants are described as follows:

$$m_\infty(V) = (1 + \exp(-(V - \theta_m)/\sigma_m))^{-1} \quad (4)$$

$$h_\infty(V) = (1 + \exp(-(V - \theta_h)/\sigma_h))^{-1} \quad (5)$$

$$n_\infty(V) = (1 + \exp(-(V - \theta_n)/\sigma_n))^{-1} \quad (6)$$

$$r_\infty(V) = (1 + \exp(-(V - \theta_r)/k_r))^{-1} \quad (7)$$

$$a_\infty(V) = (1 + \exp(-(V - \theta_{s1})/k_{s1}))^{-1} \quad (8)$$

$$s_\infty(V) = (1 + \exp(-(V - \theta_s)/k_s))^{-1} \quad (9)$$

$$\tau_n(V) = \tau_n^0 + \tau_n^1/(1 + \exp(-(V - thn)/sn)) \quad (10)$$

$$\tau_h(V) = \tau_h^0 + \tau_h^1/(1 + \exp(-(V - thh)/sh)) \quad (11)$$

Membrane capacitance $C$ is 1 pF. Maximal conductances in nS are $g_{Na} = 120$, $g_K = 30$, $g_{AHP} = 30$, $g_T = 0.5$, $g_{Ca} = 0.1$, and $g_L = 0.1$. Reversal potentials in mV are $V_{Na} = 55$, $V_K = -80$, $V_{Ca} = 120$, and $V_L = -55$. Parameters for $Ca^{2+}$-current are $k_{Ca} = 3$ and $\varepsilon = 0.0055$. Other parameters are $\theta_m = -37 \text{mV}$, $\theta_h = -58 \text{mV}$, $\theta_n = -50 \text{mV}$, $\theta_r = -70 \text{mV}$, $\theta_{s1} = -35 \text{mV}$, $\theta_s = -57 \text{mV}$, $thn = -40 \text{mV}$, $thh = -58 \text{mV}$, $\sigma_m = 10$, $\sigma_h = -12$, $\sigma_n = 14$, $k_r = -2$, $k_{s1} = 2$, $k_s = 2$, $\tau_r = 10$, $\tau_n^0 = 0.05$, $\tau_n^1 = 0.27$, $sn = -12$, $\tau_h^0 = 0.05$, $\tau_h^1 = 0.27$, $sh = -12$, $k_1 = 30$, $\phi_n = 0.3$, $\phi_h = 0.1$, $\phi_r = 1$, and $\phi_n = 0.3$.

We generated the model STN data using the STN cell model in [15,16,19]. It is described by the same equations as Eqs. (1-3), but parameter values are different [15,16,19]. The model STN generates periodic bursty input, which is delivered to the GPe cell through a model excitatory glutamatergic synapse represented by the first-order kinetic equation describing the fraction of activated channels

$$ds/dt = \alpha H_\infty(V_{presyn} - \theta_g)(1-s) - \beta s \qquad (12)$$

where the sigmoidal function $H_\infty(V) = 1/(1+\exp[-(V-\Theta_g^H)/\sigma_g^H])$. The synaptic current from the STN-to-GPe synapse is then given by $I_{syn} = g_{S\to G}(V - V_{syn})s$. Parameters for synaptic current are $\alpha = 2$, $\beta = 0.12$, $\theta_g = 20$, $\Theta_g^H = -57$, and $\sigma_g^H = 2$. $g_{S\to G}$ is set here to 0.5 unless stated otherwise.

For the comparison of the model GPe response, we also generated STN activity using spike trains, which were experimentally obtained from advanced parkinsonian patients and delivered them to the GPe cell through synaptic variable *s*. The model parameters for STN cell follows [15]. Numerical integration of the model was performed with XPPAUT (http://www.math.pitt.edu/~bard/xpp/xpp.html).

## 2. Data analysis

### *Maximal Lyapunov exponent*

To show that GPe response to a periodic rhythmic input is chaotic, we computed the maximal Lyapunov exponent using slow variable [*Ca*]. After a 5-second transient period, we used 30-second data to collect the maximum value of [*Ca*] near the end of GPe burst. During the burst of GPe cell, the overall level of [*Ca*] increases slowly. We used TISEAN [22] for computation of Lyapunov exponents. Positive maximal Lyapunov exponent is an evidence for a chaotic activity.

### *Phase and phase synchronization index γ*

Both time-series of STN and GPe cells were down-sampled at a 1 kHz rate and filtered to the spectral band of 10-30 Hz (beta band) in order to study phase synchronization within beta band. The phases of the signals were computed from the Hilbert transform. The resulting phase is the phase for bursting in the beta band, as this band is relevant to parkinsonian symptoms (see Introduction). The synchronization index is defined by

$$\gamma_N(t_k) = \left\| \frac{1}{N} \sum_{j=k-N}^{k} e^{i\Phi_j} \right\|^2 \qquad (13)$$

where *N* is the number of data points preceding time $t_k$ and $\Phi_j = \phi_{STN}(t_j) - \phi_{GPe}(t_j)$ is the difference between the phases of STN and GPe cells. In this paper, we used 512ms-long window (*N* = 512). For the confidence level of $\gamma$, we used "S4-type" surrogate data which preserve important spectral properties of the original data as in [9,10].

*First-return maps for the phases*

The temporal structure of phase synchronization between STN and GPe cells shows an intermittent pattern, that is, the synchrony is irregularly interrupted by some desynchronizing events. To characterize whether this intermittency is driven by short-duration but frequent events or sparse but long-duration events, we used a first-return map and computed rates as shown in [9,23]. Briefly, first, we set up a check-point for the phase of STN cell (0 in this study but not crucial for the analysis) and recorded the phase of GPe cell $\{\phi_{GPe,i} | i = 1,\cdots,N\}$ whenever the phase of STN cell crosses this check point from negative to positive. $\{\phi_{GPe,i}\}_i$ was translated so that mean is $\pi/2$ and we plotted $\phi_{GPe,i+1}$ vs. $\phi_{GPe,i}$ for $i = 1,\cdots,N-1$. A synchronous regime would result a cluster of points on the diagonal $\phi_{GPe,i+1} = \phi_{GPe,i}$ while completely uncorrelated phases of the signals would yield a $(\phi_{GPe,i}, \phi_{GPe,i+1})$ space homogeneously filled with dots. $(\phi_{GPe,i}, \phi_{GPe,i+1})$ space is divided into four regions labeled 1 through region 4 starting from the first quadrant in clockwise direction. The transition rates between the four regions of the map, $r_i$, are defined as the number of points in a region from which the system leaves the region, divided by the total number of points in that region. Note that the phase space $(\phi_{GPe,i}, \phi_{GPe,i+1})$ represents current phase vs. future phase, so the transitions in this space cannot be completely arbitrary. All rates vary between 0 and 1. $r_1$ tells how frequently the phase synchronization is lost and the other three rates characterize the dynamics of desynchronization events. The details of this time-series analysis approach are available in [23].

Time-series analysis was done with the custom-written programs in MATLAB (Mathworks, Natick, MA).

## Dynamics of GPe cell under STN oscillatory input

In [15], we found that the parameter region for the experimentally observed intermittent synchrony resides in the large boundary between synchronized and nonsynchronized dynamics. Over this boundary region, $I_{app}$ ranges from 4 to 10 (figure 4 in that paper). In the current study, $I_{app}$ is fixed at 7 as one representative value within that range of $I_{app}$. As $I_{app}$ increases, inter-burst interval decreases slightly and number of spikes per burst eventually increases by 1 when $I_{app}$ is 8. However, the results under different $I_{app}$ values are qualitatively the same, For $I_{app} = 7$, the model GPe model fires bursts with 3 spikes where two inter-burst intervals, 51.15 ms (std. 0.0549) and 40.92 ms (std. 0.0536), alternate (Fig.1A). As explained in Methods section, the activity of the STN cell was delivered to the GPe cell through synaptic variables to give a better approximation than using a pulse-like input. Due to the non-instantaneous nature of synaptic variable, it takes some time for firings of the STN cell to affect the GPe cell. The strength of the STN to GPe synapse $g_{S \to G}$=0.5. This is a relatively strong synaptic strength, so that the GPe cell responds faithfully to synaptic inputs in most cases.

We first check responses of the GPe cell to a single burst consisting of 3 spikes with inter-spike intervals 7.1 ms. When the input is given during the silent phase of GPe cell, it responds faithfully by firing 3 spikes and then there is long silent period until next active phase which has a larger number of spikes. Fig. 1B shows an example of 5 spikes per burst in GPe cell. In general, as the phase at which the GPe cell receives input from the STN cell increases, the duration of the silent period to the next active phase decreases and so does the number of spikes within the next burst. When the onset of stimulus approaches the onset of burst in GPe cell, the response shows greater sensitivity depending on the input phase. Fig. 1C-E illustrate these sensitivities when inputs are given at 58.9, 59.9 and 60.9 ms (denoted by stars in the figure). Depending on the input phase, a slight phase advance but relatively no change (Fig. 1C), a significant phase advance with smaller number of spikes per burst (Fig. 1D) or a considerable phase delay (Fig. 1E) are generated. Depending on the input phase, there are also differences between the numbers of spikes within next active phase after the input. For example, we see 3, 2, and 4 spikes per next duration of active phase in Fig. 1C, D and E, respectively. This sensitivity appears to be essential for the intermittent synchronization between subthalamic input and pallidal output, as considered in the next part of the paper.

Now we check the response of the GPe cell to rhythmic synaptic inputs driven by STN activity. We used a repeated 3-spike bursting pattern, a periodic input with period *T* between 40 ms and 100 ms, which may roughly reflect the beta band activity associated with the pathological state (see Introduction for references).

A rhythmic response of GPe cell to the periodic STN activity is shown in Fig. 2A. The number of spikes within a GPe cell burst alternates between 4 and 5. The first burst in the figure illustrates an elongation of the GPe cell response due to the input from STN cell, which was given in the middle of GPe cell's active phase. Additional firings are driven by STN input and the extended burst ends as the input terminates. On the other hand, the second burst in the same figure shows an initiation of GPe cell response by the STN input. In this case, although the input is terminated, GPe cell fires additional spike because the calcium level is still insufficiently high. In either case, the spikes within the burst are tightly phase-locked to those of the input.

Fig. 2B, on the other hand, demonstrates irregular activity patterns of GPe cell response to periodic STN input (period is different from those in Fig. 2A). GPe cell's standalone bursts are irregularly mixed with two types of GPe cell response bursts shown in Fig. 2A such as the elongated response bursts (the 1st and 6th bursts in the figure) and the STN input initiated burst (the 7th one). The GPe cell's spontaneous bursts demonstrate variable inter-burst intervals as well as a variable number of spikes. In most cases, GPe cell responds faithfully to the firings of STN cell even though it is given in the middle of its active phase. As apposed to Fig. 2A, however, the GPe cell also demonstrates the ability to fire an additional spike between STN input spikes although the response burst is initiated by the STN input (for example, the 2$^{nd}$ and 7$^{th}$ bursts).

For illustrative purposes, in Fig. 2C we used an experimentally obtained STN spike train as synaptic input timings in the model and considered the response of the model GPe cell (see Methods for the details of experimental data). Rhythmic firing in the beta band is clearly shown in STN spike train (black trace). The response of the model GPe cell under this input demonstrates similar patterns to those shown in Fig. 2B. For example, we can see standalone bursts with variable number of spikes, burst elongated by STN input, and bursts initiated by STN input.

To see if the GPe response under periodic bursting input is chaotic, we computed maximal Lyapunov exponents (MLEs) using data obtained through Poincare section. We chose intracellular calcium concentration [$Ca$] for this purpose. It is reasonable choice because, as will be explained later, activity patterns of the GPe cell are closely related to and can be described by [$Ca$]. For example, overall level of [$Ca$] increases over the active phase and monotonically decreases over the silent phase. We specify that GPe cell enters silent phase if [$Ca$] assumes its maximum over active phase. Whenever the GPe cell finishes firing and enters silent phase, we recorded values of [$Ca$]. The collection of these values, $\{[Ca]_n\}$, is used for computation of MLEs. Resulting MLEs were 0.577 (Fig. 2B) and 0.4356 (Fig. 2C). Positive MLE indicates that the activity pattern is chaotic and resulting intermittent synchrony is really irregular. While real neurons may have substantial stochastic components in their dynamics, this suggests that some degree of irregularity of synchrony may be induced by purely dynamical factors. Moreover, as we will show below, small stochasticity added in the model does not destroy the observed phenomena.

Response of the GPe cell to periodic bursting input also shows sensitivity to the input period, $T$. Figure 3 shows exemplary GPe cell responses for $T$ = 58, 59, 60, and 61 ms. We used a 25-second long data-window after a 5-second long transient period to collect $\{[Ca]_n\}$ and plotted ($[Ca]_n$, $[Ca]_{n+1}$). This is a descriptive way of illustrating GPe cell responses. If the GPe cell response is periodic, then only few points are shown. Periodic responses when $T$ =59 ms and 60 ms are clearly shown in the figure in contrast to irregular responses when $T$ = 58 and 61 ms.

## Geometric analysis of periodic response of GPe cell

If the GPe cell shows periodic response, then we can derive an analytic map using standard fast-slow analysis. A characteristic feature of fast-slow analysis is that one can describe dynamics of the full system with dynamics of the slow variables. Numerical simulations indicate that the intracellular calcium concentration, [$Ca$], is slower than the other dependent variables, so we treat [$Ca$] as the slow variable and the others as fast variables in our fast-slow analysis. To begin with, we consider [$Ca$] as bifurcation parameter in the fast subsystem, which is a system of equations of fast variables. Figure 4A shows the resulting bifurcation diagram of the fast subsystem. There is a unique fixed point for each value of $[Ca]$. A horizontal line in the middle of the figure denotes the set of these fixed points. On this line, there is a special point called a subcritical Hopf bifurcation point when $[Ca] = [Ca]_{HB}$, from which branches of an unstable periodic orbit emanates. Fixed points are stable for $[Ca] > [Ca]_{HB}$ (thick line) and unstable for $[Ca] < [Ca]_{HB}$ (thin line). We have another special point called a saddle-node of periodic orbits when $[Ca] = [Ca]_{SN}$, at which these branches of an unstable periodic orbit turn around to be branches of a stable periodic orbit. Thus, for $[Ca]_{SN} < [Ca] < [Ca]_{HB}$, the fast system shows bistability (stable fixed point and stable periodic orbit).

Next we include the dynamics of a slow variable $[Ca]$ when the GPe cell shows bursting. Figure 4 B shows the projection of a spontaneous bursting solution (Figure 1A) onto the bifurcation diagram. We trace the bursting solution from the point when the trajectory jumps down from its active phase and begins the silent phase (circle). At this time, $[Ca]$ assumes its maximum. During the silent phase, the trajectory approaches the branch of stable fixed points of the fast subsystem while $[Ca]$ decreases monotonically. Thus, the trajectory moves leftward close to the branch of stable fixed points until it reaches the Hopf bifurcation point. If it passed the Hopf point, then it eventually jumps up into the branch of stable periodic orbits and generates action potentials (square). Over the active phase of the cell, the overall level of $[Ca]$ increases. Once it reaches the saddle-node of limit cycles, then the trajectory jumps down and this completes one cycle of the bursting solution.

There is a delay from when the trajectory passes the Hopf point to when it jumps up (a delayed bifurcation). This is not obvious in Figure 4B but is clearly illustrated in Figure 4C, which shows the projection of a bursting solution with input from the STN cell (Figure 1C). The middle gray trace is the last action potential of spontaneous bursting before STN input, which is followed by three STN input-driven action potentials (black traces on the right). Once the trajectory jumps down, then it tracks close to the branch of fixed points until it jumps up to generate action potential (leftmost black trace). There is a significant delay in this case as compared to Figure 4B.

In fact, previous analysis has demonstrated that the length of the delay is closely related to the distance between the Hopf point and where the cell jumps down. We numerically computed this relation and the results are shown in Figure 4D. The horizontal axis represents the $[Ca]$ value when the cell jumps down and the vertical axis the $[Ca]$ value when the cell finishes decreasing and jumps up into active phase. Numerical results are shown in gray and linear fitting is in black. In fact, this linear fitting is sufficient for finding analytic maps, so we use it in the following analysis. The equation of linear fitting is given by $y = -0.4x + 0.44$.

Now we are ready to derive analytic maps using averaged dynamics of slow variable $[Ca]$. For clarity of explanation, we repeat the governing equation of $[Ca]$ (Eq. 2) here.

$$\frac{d[Ca]}{dt} = \varepsilon(-I_{Ca} - I_T - k_{Ca}[Ca]) \qquad (2)$$

where $\varepsilon = 0.0055$ and $k_{Ca} = 3$. We first averaged the quantity, $-I_{Ca} - I_T$, in the governing equation of $[Ca]$ over the active and silent phases respectively and got 2.63 and 0.09, respectively. The resulting equations are given as

$$\frac{d[Ca]}{dt} = \begin{cases} \varepsilon(2.63 - k_{Ca}[Ca]) & \text{when the cell is in active phase} \\ \varepsilon(0.09 - k_{Ca}[Ca]) & \text{when the cell is in silent phase} \end{cases} \qquad (14)$$

Solutions with initial condition $C_0$ are given by $[Ca](t) = 0.877 + (C_0 - 0.877)\exp(-3*\varepsilon*t)$ in active phase and $[Ca](t) = 0.03 + (C_0 - 0.03)\exp(-3*\varepsilon*t)$ in silent phase.

We fixed input period $T$ and assumed that STN input begins at time $t = 0$ and lasts 15ms. We also assume that GPe cell enters silent phase as soon as STN input duration ends. There remains $(T-15)$ ms for the next STN input. Now we will trace values of $[Ca]$ whenever the cell enters silent phase. We begin with assuming that $[Ca] = C_0$ at $t = 15$ms. Without any influence from the STN cell, the lowest value that $[Ca]$ can reach is given as $C_L(C_0) = -0.4C_0 + 0.44$ (Figure 4D). On the other hand, the value of $[Ca]$ after $(T-15)$ ms while the cell is in silent phase is given as $C_S(C_0) = 0.03 + (C_0 - 0.03)\exp(-3*\varepsilon*(T-15))$. Let $C_0^*$ be the value of $[Ca]$ such that $C_L = C_S$, which is given as

$$C_0^* = \frac{0.41 + 0.03\exp(-3*\varepsilon*(T-15))}{0.4 + \exp(-3*\varepsilon*(T-15))} \qquad (15)$$

If $C_0 > C_0^*$, then we have $C_L < C_S$, so the cell gets input from STN cell before it reaches $C_L$ and jumps into active phase. At this time, the level of $[Ca]$ is given by $C_1 = 0.03 + (C_0 - 0.03)\exp(-3*\varepsilon*(T-15))$. With $C_1$ as an initial condition, the cell proceeds for 15ms while it gets input from STN cell. Thus $[Ca]$ value when the cell enters silent phase again is given as $0.877 + (C_1 - 0.877)\exp(-3*\varepsilon*15)$. If $C_0 < C_0^*$, on the other hand, then $C_L > C_S$, so the GPe cell reaches $C_L$ and enters active phase spontaneously. Let $T^* = T^*(C_0)$ be the time at which the cell reaches $C_L$, then is given as

$$T^* = 15 - \frac{1}{3\varepsilon}\ln\left(\frac{-0.4C_0 + 0.41}{C_0 - 0.03}\right) \qquad (16)$$

by solving $-0.4C_0 + 0.44 = 0.03 + (C_0 - 0.03)\exp(-3*\varepsilon*(T^* - 15))$. At $t = T^*$, the cell enters active phase with $C_L$ as an initial condition. GPe cell proceeds for $(T - T^*)$ ms until it gets the next input from STN cell and for additional 15ms during STN input. Thus the value of $[Ca]$ when the cell enters silent phase is given as $0.877 + (C_L - 0.877)\exp(-3*\varepsilon*(T - T^* + 15))$. In summary, we derived the following map

$$F(C_0) = \begin{cases} 0.877 + (C_1 - 0.877)\exp(-3*\varepsilon*15) & \text{if } C_0 > C_0^* \\ 0.877 + (C_L - 0.877)\exp(-3*\varepsilon*(T - T^* + 15)) & \text{if } C_0 < C_0^* \end{cases} \quad (17)$$

where $C_1 = 0.03 + (C_0 - 0.03)\exp(-3*\varepsilon*(T - 15))$ and $C_L = -0.4C_0 + 0.44$.

Examples of maps when $T = 58, 59, 60$, and 61 ms are shown in Figure 5 (gray dots). Numerically computed results in Figure 3 are denoted by black squares in Figure 5. We also plotted iteration of each map after some transient period (gray diamonds). Note that only two points were generated from iteration of each map; this fact implies that "theoretically" GPe cell responses are periodic with period 2.

When the response of the GPe cell to rhythmic STN inputs is periodic, maps from geometric analysis give good approximation to numerical results (Fig. 5B and 5C). Naturally, when the numerical GPe cell response is really periodic with period 2, the theoretical result provides good matching to numerical results (Fig.5C). Fig. 5B illustrates another example where GPe cell response is periodic with longer period. In this case, two points from numerical simulation lie close to points from maps while there are another two points from numerical simulation away from these points. When the GPe cell response is not periodic, we see that points from numerical results were scattered around the analytical map (Fig. 5A and 5D).

Thus the response of the GPe cell to rhythmic STN input may tend to be periodic with period 2. But this periodic response of the GPe cell is disrupted or eliminated due to the sensitivity of the GPe cell response to arrival timing of STN input as shown in Figure 1. How often sensitive response of the GPe cell occurs depends critically on input period $T$. Depending on input period $T$, periodic response of GPe cell is preserved, disrupted slightly, or disrupted severely to become aperiodic eventually.

When GPe cell shows aperiodic responses, the synchrony between STN and GPe cells demonstrates similarity to experimentally observed intermittent phase-locking. In the following section, we will first look into the temporal structure of synchrony between STN and GPe cells (to see how it compares with experimental observations) and then study how it also depends on other properties of incoming STN input defined by cellular properties of STN cells which are affected by dopamine loss.

**Relationship between phases of GPe cell and STN cell**

We now consider the details of how GPe activity is correlated to a periodic STN input in the case when GPe response is aperiodic ($g_{S \to G}$ = 0.5 and input period $T$ = 85 ms, Fig. 2B). To study the temporal structure of synchrony between the input and the output, we computed time dependent phase locking index γ (see methods.). Fig. 6A shows the resulting dependence of γ on time (black line) against the value of γ obtained from surrogate data (gray line). Similar to what we found in experimental data [9] and our previous modeling study [15], the fine temporal structure of the synchrony shows intermittent characteristics; the synchrony between them fluctuates frequently, in other words, it is interrupted by desynchronizing events. A model GPe cell can generate irregular responses to periodic input, which can be a potential contribution to intermittent synchronization within the STN-GPe network.

We also considered the response of the GPe cell under noisy STN input to confirm the robustness of the phenomena we study. To do this, we kept input period almost constant but jittered STN spikes within the STN burst slightly. We achieved this by adding noise to the equation for the voltage of STN cell over STN bursting duration. We used a Wiener process (a standard algorithm built in XPPAUT) to generate noise and checked a range of amplitudes from 0.1 to 1. Addition of noise resulted in shifts of STN spikes and additional spikes sometimes. While an increased level of noise lowered the overall level of synchrony level, γ, we found that the observed characteristic intermittency of synchrony is persistent over noisy STN input.

Now let us explore the fine temporal structure of this synchronization [9,23]. The same moderate phase-locking level may be achieved with frequent short deviations from the phase-locked state, with a few long desynchronization episodes, and with a variety of options in between. To see how often the synchrony between STN and GPe is interrupted by desynchronizing events and how quickly the synchrony is restored, we checked first-return maps using phases of the GPe cell when $g_{S \to G}$ = 0.5 and $T$ = 85 (see Methods). Figure 6B shows the resulting first-return map with significant two branches, which resembles those observed in experimental and modeling studies [9,15]. For comparison, we provide an example of first-return map from experimental data, recorded in parkinsonian basal ganglia, that is from the phase of the local field potential, LFP (Fig. 6C). Note that STN LFP is expected to be mostly generated by pallidal synaptic input to STN and is, therefore, representative of GPe activity (see discussions in [9,15]). The cluster on the diagonal line reflects the tendency for phase entrainment between the input and the output while two branches represent desynchronization events. The rate $r_1$, the rate of deviation from the "phase-locked state" (defined here broadly, as the state with phase difference less then π/2, see Methods), is 0.2264 in Fig. 6B and 0.2604 in Fig. 6C.

To check how a long desynchronized state persists once the synchrony is broken, we recorded the durations of those desynchronization episodes (Fig. 6D). Cycle length 1 means that $\phi_n$ deviates from the "synchronized state" for one cycle of oscillations, cycle length 2 is defined similarly etc. The histograms of model system and experimental data have visible differences (e.g., desynchronizations lasting for 3 cycles of oscillations is more prevalent in the model, rather than in the experimental data). However, in both cases the shortest desynchronizations prevail.

## Dependence of synchronous dynamics on dopamine-modulated properties of STN firing.

The results above indicate that, for some specific set of parameter values, the response of the GPe cell to rhythmic STN input can generate a chaotic response and the synchrony between the input and the output retains the important characteristics observed in experimental data. As we described above, Parkinson's disease and its hypokinetic motor symptoms are characterized by the loss of basal ganglia dopamine and resulting synchronized patterns of neural activity. Therefore we consider how the synchrony between STN input and GPe output changes as we vary dopamine-modulated parameters. This would show how dopamine modulation of cellular properties influences the dynamics of our model system.

Earlier computational studies explored the effect of dopamine-modulated synaptic strength on synchronization in the STN-GPe network [15,16,18,19]. However, experimental results have demonstrated that dopamine also affects the membrane properties of the neurons. This naturally should lead to the changes in timing of neural discharge. In particular, the STN cell appears to be depolarized by the action of dopamine, so that if it is bursty, the duration of the burst, the duration of the inter-burst interval and the number of spikes in the burst are increased while the frequency of spikes per burst decreases [21,24,25,26]. In that sense, STN cell can be considered as being less bursty because the distinction between the duration of active phase of bursting STN cell and inter-burst interval weakens. So we will focus here on the effect of dopamine on the cellular properties of STN neurons (and this STN discharge pattern) in the generation of intermittent synchronization between STN and GPe.

Following the aforementioned experimental results, we consider three parameters, which should be affected by dopamine. These are the duration of STN active phase ($T_b$), the period of STN input ($T$) and the inter-spike interval (*ISI*) within the active phase of STN input to GPe. The number of spikes per burst then depends on $T_b$ and *ISI*. According to experimental results, $T_b$, $T$ and *ISI* would increase as dopamine level increases. We also expect that the ratio $T_b/T$ increases because STN cell becomes less bursty.

We first fix the inter-spike intervals of STN input to GPe at multiple values and check the level of synchrony between the input and output over the parameter space of $T_b$ and $T$. This creates several section planes in the 3-dimensional parameter space. To measure the synchrony between the input and output, we divide the time series of membrane potentials into non-overlapping 1s-long windows, compute phase synchronization index γ (Eq. (13)) over each window and average the results. Figure 7 shows the results for several values of *ISI*. As *ISI* decreases, the region of higher synchrony broadens from upper left corner to lower right corner. Figure 7 demonstrates that higher synchrony between the input and the output can be obtained when we have larger $T_b$ and smaller $T$, so $T_b/T$ is larger. In other words, if the input frequency is relatively high and the proportion of active phase duration per input period is large, then we would get synchronization between the input and the output. In addition, if firing frequency within a burst of STN input increases, then the minimum value of $T_b/T$ required to get synchrony can be lowered.

These results look reasonable because we have sufficiently strong synaptic connections from the STN cell to the GPe cell for the GPe cell to respond faithfully to the input from the STN cell. As seen in Fig. 2, the GPe cell may generate additional spikes during the input burst or between the input bursts. However, the chances of this happening can be reduced by frequent inputs and a longer duration of active phase per input period as shown in Fig. 2A. In other words, due to the frequent inputs, the GPe cell cannot fire between inputs and tends to be phase-locked to STN input. This synchronization can be more effectively obtained if we have a larger number of spikes per active phase of STN cell. Smaller *ISI* further reduces the chance of additional spiking during STN input as shown in Fig 2B.

Thus, a transition from synchronized state to nonsynchronous state can be observed by changing two parameters, decreasing $T_b$ and increasing $T$. These changes, however, are not necessarily relevant to the experimental results, because in experiments [21,26] both $T_b$ and $T$ increase as dopamine level increases. Thus, although this result is intuitively reasonable, it does not give much insight into the actual mechanism underlying transition from pathological state to normal state. Hence, this result may suggest that we need an additional parameter to achieve the transition.

Therefore we examine how the synchrony level changes along different paths in the three-dimensional parameter space. We chose paths along which the number of spikes per input burst is fixed. Recall that the number of spikes per burst is determined by *ISI* and $T_b$. For a fixed number of spikes per burst, *ISI* is roughly proportional to $T_b$. Figure 8A, B, C shows the results along some paths which have 2, 3, 4 spikes per input burst respectively. For the values of (*ISI*, $T_b$) for the individual trajectories, refer to the caption of Figure 8. Synchrony levels measured by averaged $\gamma$ show overall dependence on $T$ along the trajectories. As $T$ increases, the synchrony index $\gamma$ decreases while there are relatively large disparities between the trajectories when the number of spikes per burst is 2 (Fig. 8A). These differences between trajectories are significantly reduced as the number of spikes per input burst increases, (Fig. 8B, C). Interestingly, when the number of spikes is 3 or 4, synchrony between the input and the output doesn't show strong dependence on the ratio $T_b/T$.

Figure 8 suggests that we might get a 2-dimensional surface in 3-dimensional parameter space which shows a transition from low coherence state to high coherence state. Figure 9A provides one example of such a surface projected onto ($ISI$, $T_b$)-plane. To get a transition from high coherence state (lower right region) to low coherence state (upper left region), we should increase $ISI$ and $T_b$. For each ($ISI$, $T_b$) value, we have a different $T$ value. In general, $T$ decreases monotonically along the horizontal or vertical lines in ($ISI$, $T_b$)-plane. Over the lower synchrony area $T_b/T$ is around 0.3; it is around 0.15 over the higher synchrony region. Along the diagonal band from lower right region to upper left region, we have overall increase of $T_b/T$. We chose one route from the upper left to lower right area along the diagonal band and plotted the corresponding averaged phase synchrony index $\gamma$ in Figure 9B. We can see a relatively smooth transition from lower to higher synchrony level. Along that trajectory, the number of spikes per burst increases from 2 to 3 and the mean $T_b/T$ is 0.3 over the first 5 points and 0.216 over the last 5 points.

The changes of parameter values used here to obtain the transition from a high coherence state to lower coherence are consistent with experimental results and with the overall understanding of synchrony in Parkinson's disease. Recall that to get such a transition, an increase of dopamine level is expected. Experimental results tell us that the increase of dopamine level causes the increase of $T_b$, $T$ and $T_b/T$ as well as the increase of $ISI$ within the active phase of STN input [21,24,25,26]. This modeling result shows that dopaminergic modulation of the cellular properties of the STN cell has the ability to result in a transition from synchrony to nonsynchrony, as one observes during the change of dopamine level in experiment.

## Discussion

### *Summary of the results*

Activity patterns in parkinsonian brain are characterized by intermittently synchronized oscillations within basal ganglia. Experimental and numerical studies have implicated that the subthalamic nucleus - external Globus Pallidus network may provide neural substrate for these characteristic oscillations within parkinsonian BG. In this paper, we considered the response of a GPe cell to periodic bursting input from an STN cell within the beta band and examined the synchrony level between the input and the output. Numerical simulations demonstrate that GPe response to rhythmic STN input may be one of the mechanisms underlying intermittently synchronized activity patterns.

The GPe cell showed sensitive responses to STN input depending on the arrival phases of such input (Figure 1). Without a sensitive response of the GPe cell, results from geometric analysis showed that GPe cell response tends to be periodic with period 2. But the tendency for periodic response was disrupted slightly or severely under the presence of a sensitive GPe cell response.

Aperiodic responses may demonstrate intermittent phase-locking between STN and GPe activity patterns, which was also observed in experiments. Since aperiodic responses were due to the sensitive response of the GPe cell (defined by GPe cell's intrinsic properties) intermittent synchrony may be attributed to cellular properties of GPe cell. Note that it is not completely certain that the current GPe cell model adequately and completely captures underlying cellular physiology. However, it has some generic characteristics (such as multiple time-scales) generated by experimentally observed membrane currents. Thus results from the current study provide one potential mechanism for explaining the intermittency observed in experiments with parkinsonian basal ganglia.

Beta band rhythmic STN input resulted in either phase-locked or irregular response of GPe cell depending on three parameters, the duration of STN active phase ($T_b$), the period of STN input ($T$), and the inter-spike interval (*ISI*) in STN input burst. These characteristics of STN discharge are modulated by dopamine and thus are expected to be different between healthy and parkinsonian state. In case of irregular responses, the temporal structure of synchrony between the input and the output showed intermittent patterns (phase-locked episodes are frequently and irregularly interrupted by desynchronizing events) similar to those observed in experimental data.

As implicated by our computational results, modulation of one or two parameters was not sufficient to get the transition from a high synchrony state to a low synchrony state. For example, when *ISI* is fixed, the parameter space explored can be roughly divided into higher and lower synchrony regions (upper left and lower right corner in Fig. 7). Larger duration of active phase and smaller input period of STN discharge tend to produce higher synchrony values where their ratio is mostly greater than 0.3. As interspike interval within STN burst decreases, the region of higher synchrony broadens and the lower bound of the aforementioned ratio decreases. Although there is a transition from higher to lower coherence regions, the corresponding change of parameters is not necessarily consistent with experimental results. This may imply the need of an additional parameter to vary in numerical experiments, and indeed, variation of three parameters yields more experimentally realistic results.

Modulation of three parameters suggested by experimental results was needed to accomplish the desired transition between states. The change of synchrony level was considered along paths in the parameter space over which the number of spikes per input burst is fixed. The synchrony level tends to decrease overall as STN period increases along the path. When the number of spikes per burst is 2, there is a large variation between trajectories, which is much reduced when the number of spikes per burst is 3 or 4. A larger number of spikes per burst tend to yield more stable results. In addition, we found a 2-dimensional surface in parameter space, which shows gradual transition from lower to higher coherence. The change of parameters along this trajectory is also consistent with experimental results; as dopamine level increases, we have an increase of the duration of STN active phase and of the period of STN discharge and input ($T$), as well as a decrease of interspike interval within STN burst. These results suggest that GPe has the ability to generate intermittent synchrony under the beta band rhythmic input from STN as dopamine level controls the cellular properties of the STN cell.

*Neurophysiological implications*

As we described above, synchronized oscillations in the beta frequency band are believed to play a crucial role in the pathophysiology of parkinsonian symptoms. Modeling studies [15] indicate that as the strength of dopamine-modulated synapses varies, the model network of pallido-subthalamic neurons exhibits patterns of intermittent synchrony, reminiscent of the ones observed experimentally in Parkinsonian patients [9]. However, dopamine affects not only the strength of synaptic connections in the basal ganglia, it also affects the properties of neuronal membranes. The loss of dopamine (seen in Parkinson's disease) modifies membrane properties of basal ganglia neurons, including STN cells, resulting in the change of STN firing patterns. *In vitro* experiments have shown that STN cells are depolarized by dopamine. If the STN cell is bursty, dopamine results in longer burst duration, longer inter-burst duration, and larger number of spikes per burst with a lower frequency of spikes in the burst.

To address these dopamine-modulated changes in STN-GPe network, we chose parameters of STN discharge, affected by dopamine. As we vary them in the way one expects them to be varied in a transition from healthy to parkinsonian state, we observe transition from a lower to a higher coherent state, roughly matching the synchrony levels observed in basal ganglia in normal and parkinsonian states. The intermittent nature of the neural beta band synchrony in Parkinson's disease is achieved in the model due to the interplay of the timing of STN input to pallidum and pallidal neuronal dynamics, resulting in sensitivity of pallidal output to the phase of the arriving STN input. Thus the mechanism considered here (the change in firing pattern of subthalamic neurons through the dopamine-induced change of membrane properties) may be one of the potential mechanisms responsible for the generation of intermittent synchronization observed in Parkinson's disease.

# Acknowledgements

We thank Dr. R.M. Worth for his comments on the manuscript.

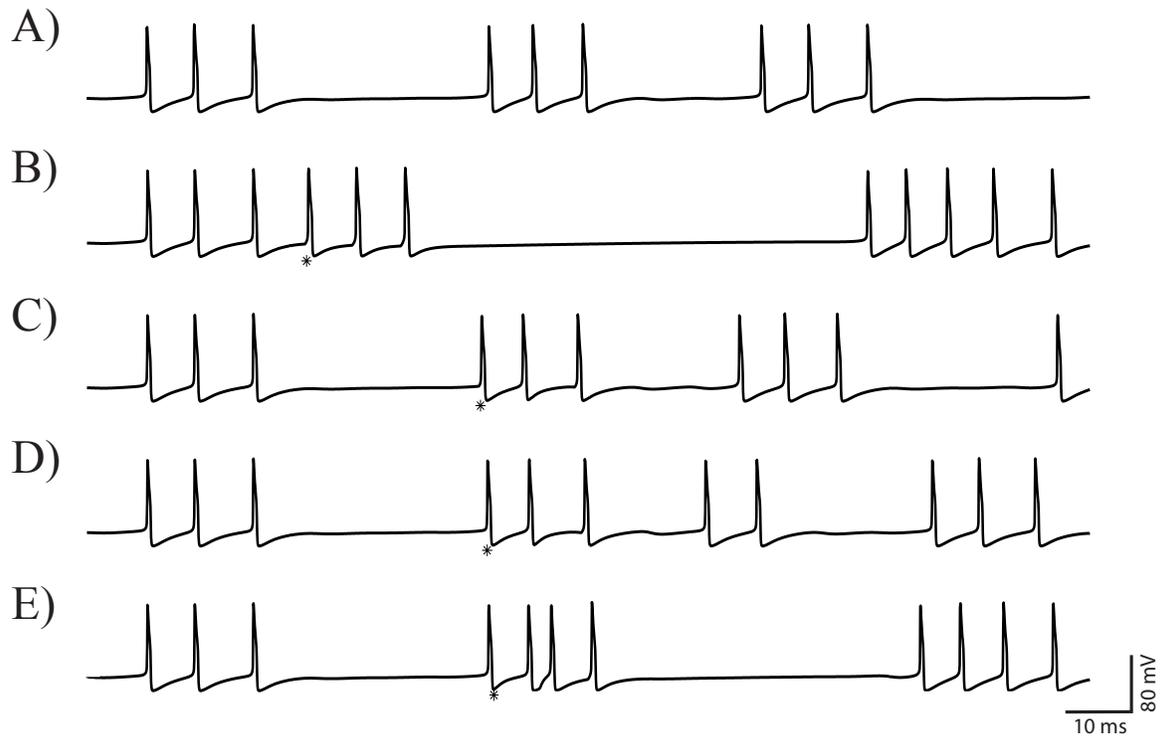

**Figure 1.** Response of a GPe cell to a single 3-spike input from STN cell. Stars denote the times when inputs are given (more precisely, the time of the peak of the first spike of the incoming burst). A) spontaneous GPe activity. B) The input during the silent phase of GPe cell causes significant phase delay. C), D) and E) show the phase sensitivity of GPe cell when the inputs are given near spiking. Depending on the exact timings, phase advance or delay are produced.

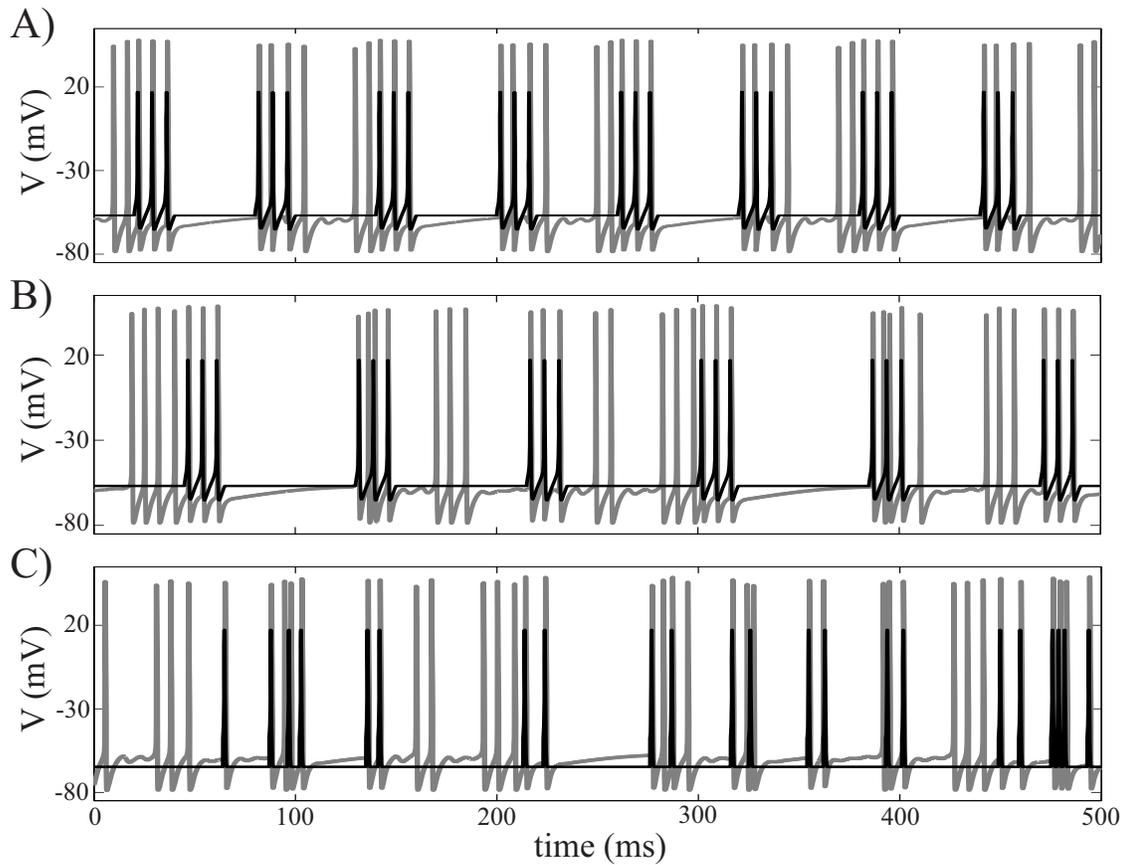

**Figure 2.** Examples of the GPe response (gray line) to rhythmic synaptic inputs driven by STN activity (black line) when $g_{S \to G} = 0.5$. A) and B) use periodic inputs generated by model STN neuron. C) uses a spike train from STN of a Parkinsonian patient recorded during surgical procedure. A) presents periodic (entrained) response when input period $T = 60$ ms and B) presents an irregular response when input period $T = 85$ ms.

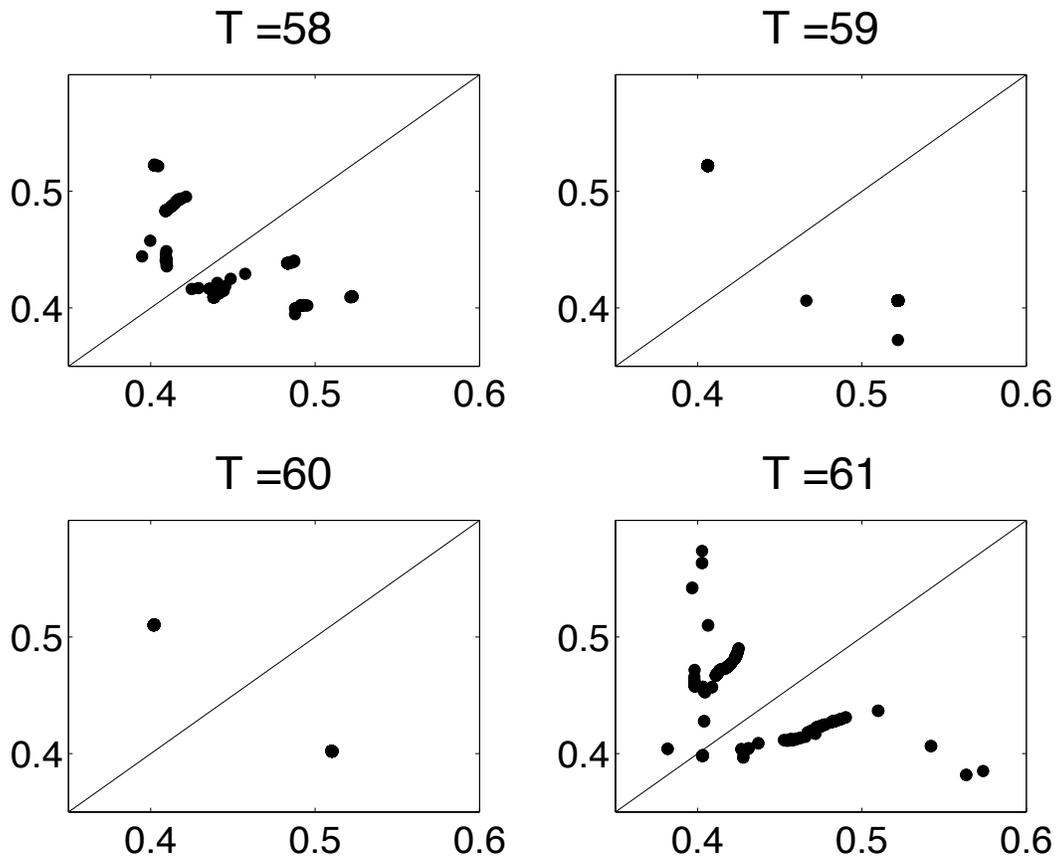

**Figure 3** Examples of the GPe cell's responses to rhythmic STN input shows significant sensitivity to the input period $T$. To describe the response of the GPe cell, we recorded $[Ca]$ values whenever the GPe cell enters the silent phase and constructed return maps. Four examples when $T = 58, 59, 60$, and $61$ ms are shown. The GPe cell shows periodic response when $T = 59$ and 60ms, which are easily disrupted by slight change of input period $T$.

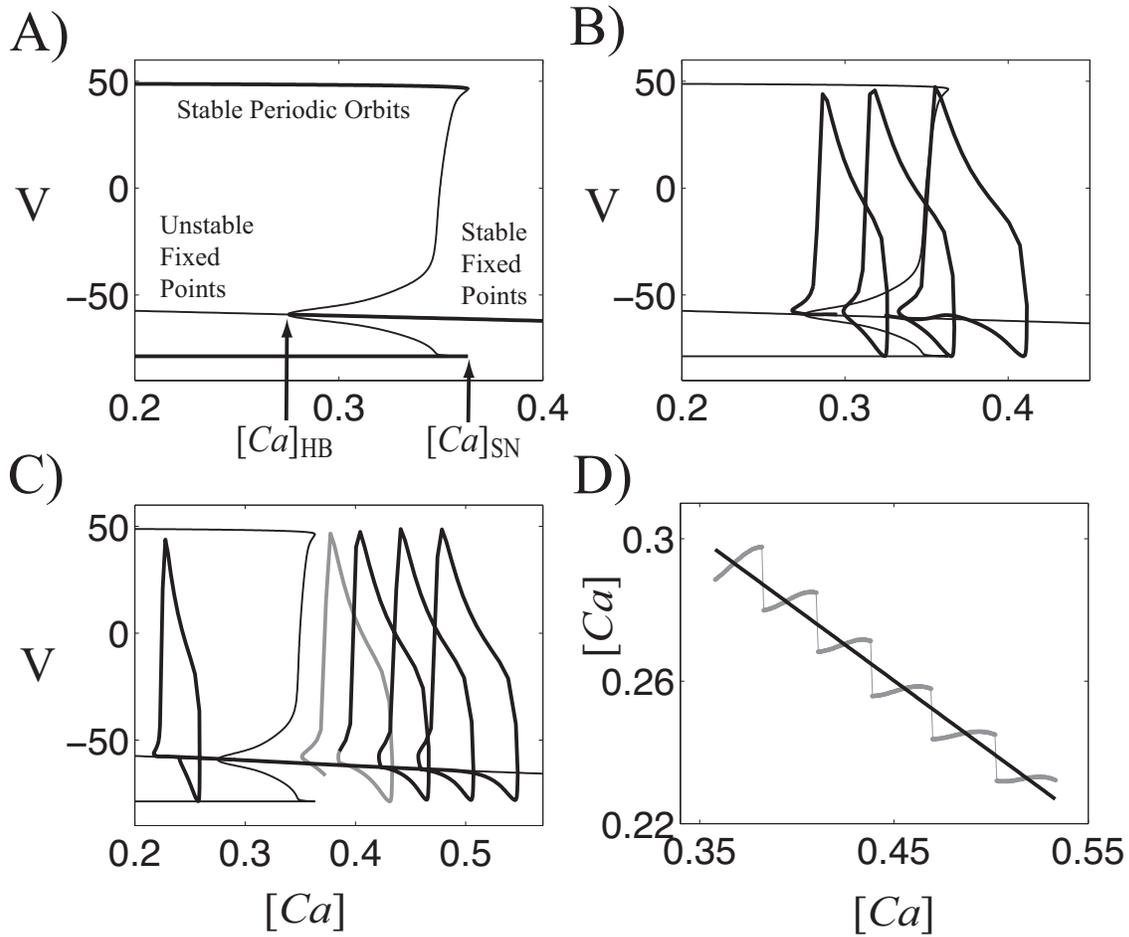

**Figure 4** A) Bifurcation diagram of the GPe cell with $[Ca]$ as bifurcation parameter. This is an elliptic burster. B) Projection of a bursting solution onto this diagram. C) An example of delayed bifurcation. D) $[Ca]$ value when it jumps up as a function of $[Ca]$ when it jumps down (gray). Black line is a linear fitting of this function.

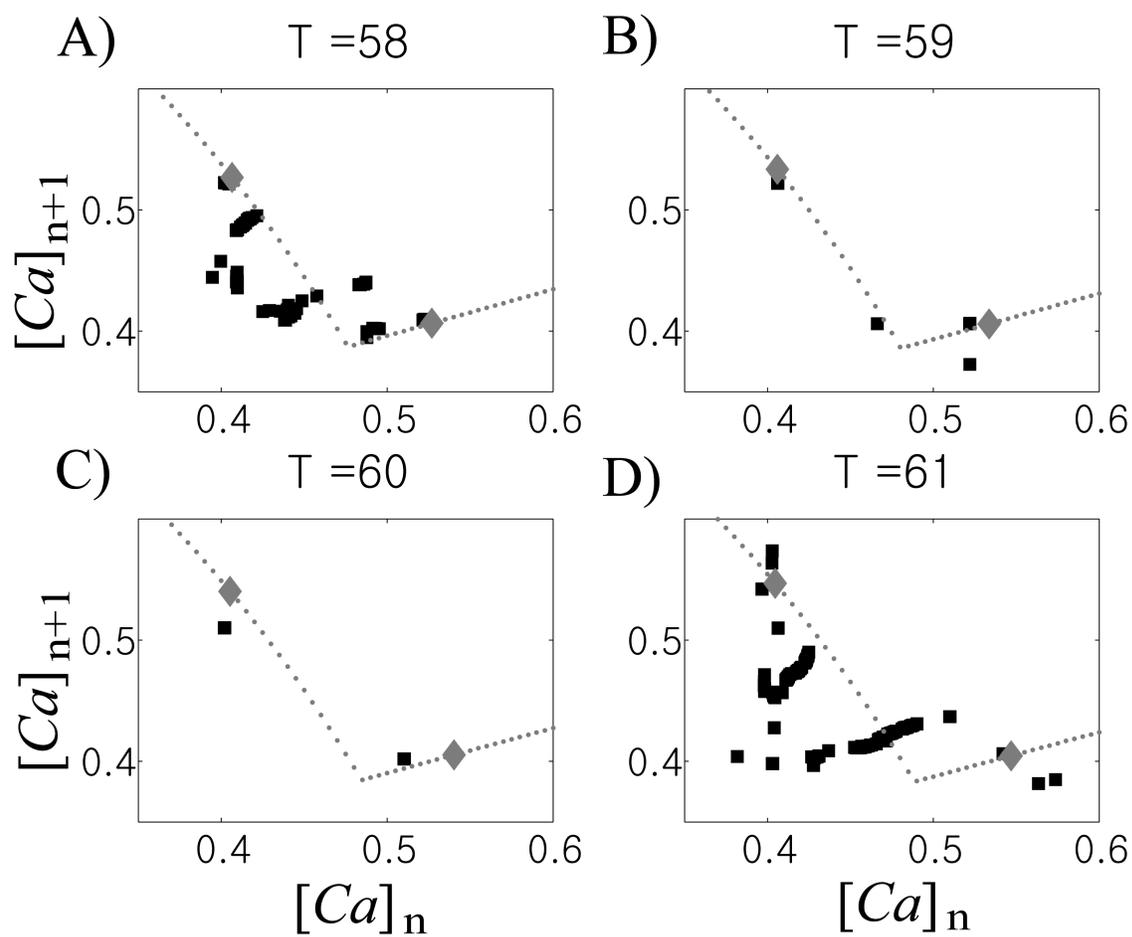

**Figure 5** Examples of first-return maps for calcium levels when *T* = 58,59,60, and 61 ms (gray dots). Squares denote numerically computed values from Figure 3 and diamonds from iteration of maps.

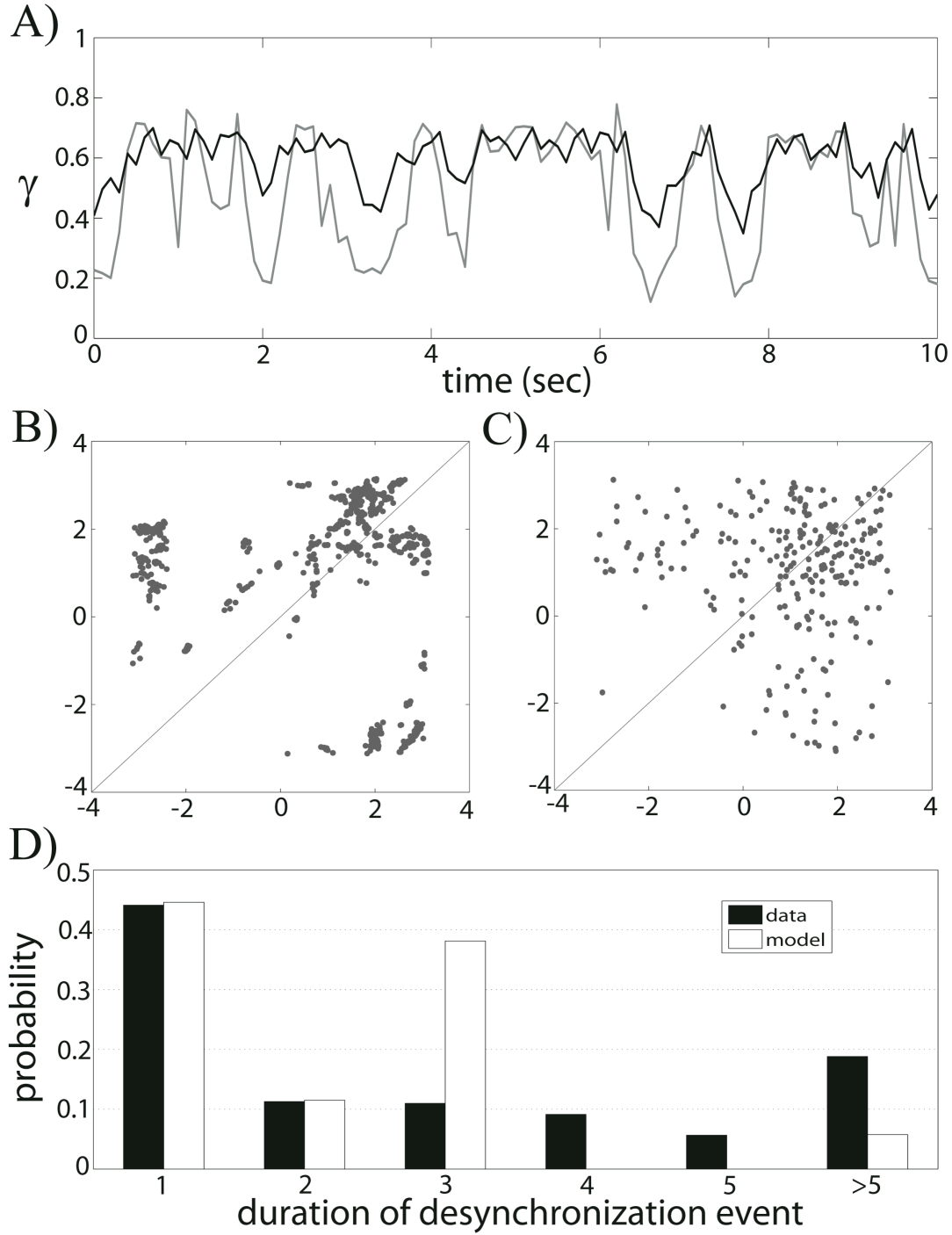

**Figure 6.** A) Temporal evolution of synchrony index γ, the window size used to compute γ is 512ms, this window precedes a point on the graph, gray line is obtained from surrogates for the confidence level of 95%. B) First-return map using the phase of GPe cell using band-pass filtered membrane potential. C) First-return map from experimental data. D) Histogram of durations of desynchronization episodes.

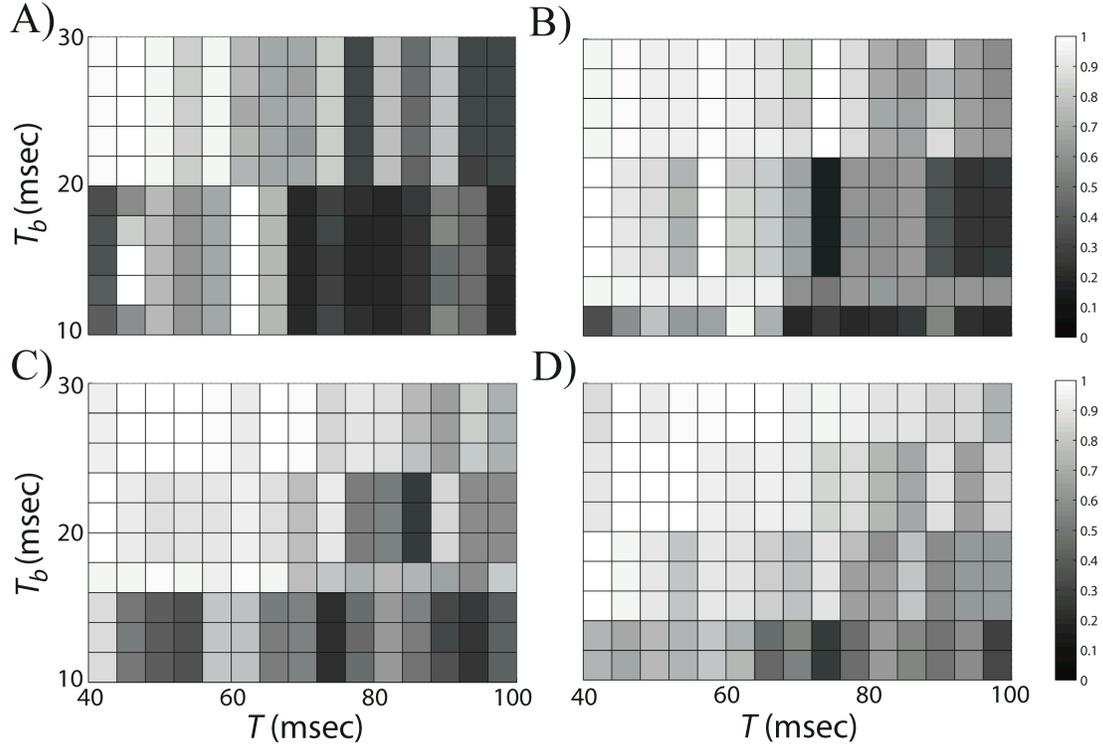

**Figure 7**. Averaged phase-locking index γ (coded by the intensity of grey tone) between the GPe cell and the STN cell in dependence on three dopamine-modulated parameters: the duration of STN active phase ($T_b$), the period of STN input ($T$) and the inter-spike interval (*ISI*) within the active phase of STN input. In each subplot, inter-spike interval within a burst of STN cell is fixed at 13ms (A), 9.2ms (B), 7.2ms (C), and 6ms (D). Number of spikes per STN input burst varies depending on *ISI* and $T_b$.

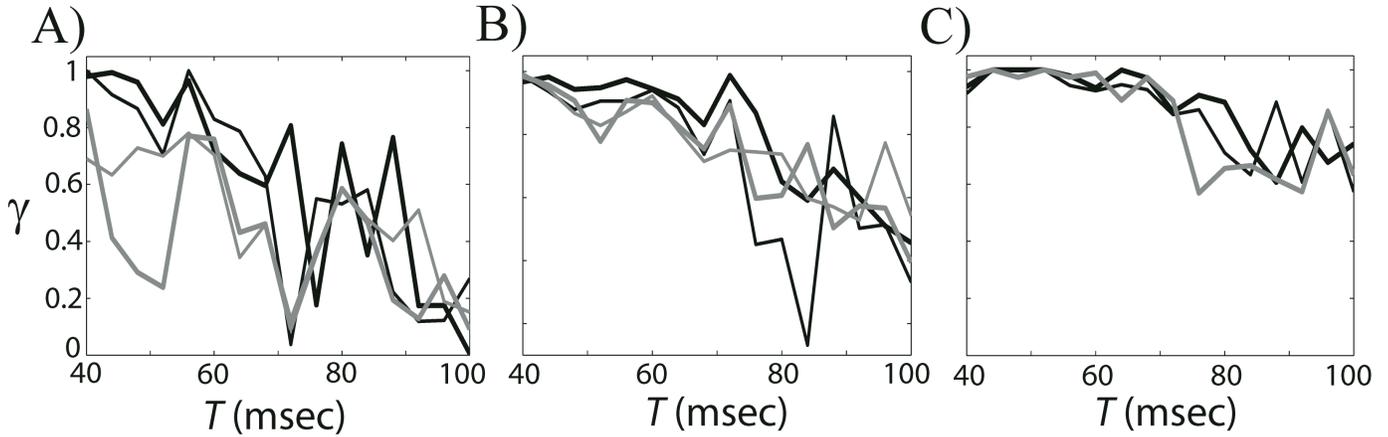

**Figure 8.** Averaged γ vs. period of STN input to GPe $T$. In each subfigure, the number of spikes in this input is fixed: 2 (left), 3 (middle), and 4 (right). The values of the inter-spike interval within the active phase and the duration of STN active phase, ($ISI$, $T_b$), are: A) thick black (13,26), thin black (9.2, 18), thick gray (7.2, 14), and thin gray (6, 10). B) thick black (9.2, 26), thin black (7.2,20), thick gray (6, 16), and thin gray (5.3, 14). C) thick black (7.2, 28), thin black (6,24) and thick gray (5.3, 20).

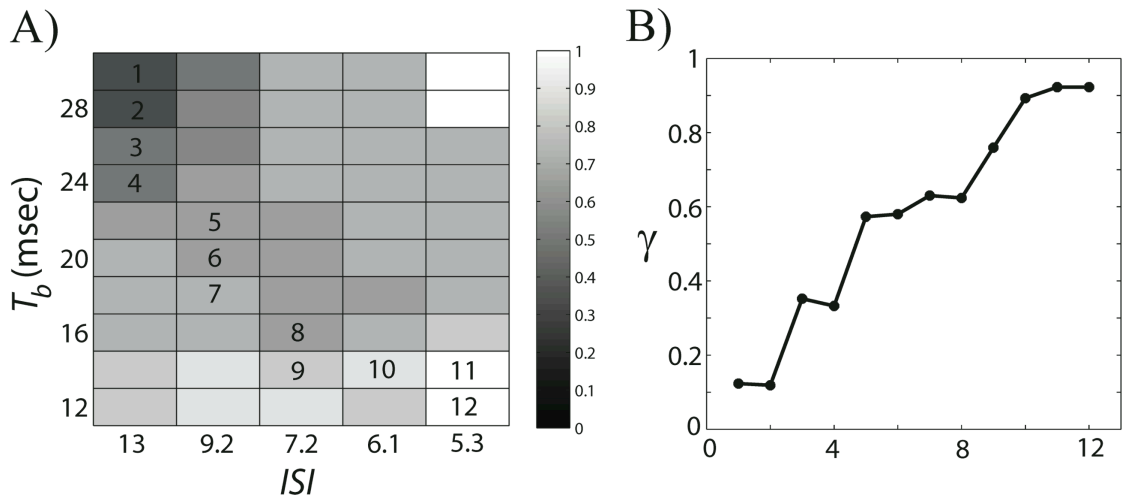

**Figure 9.** A) An example of projection of a 2-dimensional surface in a parameter space onto $T_b$-$ISI$ plane, which shows a transition from weakly coherent to stronger coherent states. Right bar shows the level of phase-locking index $\gamma$. B) We chose one trajectory from upper left corner (low coherence) to lower right corner (high coherence) to illustrate coherence transition. Numbers (1 to 12) within (A) represent the labels of the points chosen. These numbers are used in a horizontal axis in (B).